\documentclass[prl,superscriptaddress,twocolumn,longbibliography,amsmath,amssymb]{revtex4} 
\usepackage[dvips]{graphicx}
\usepackage{xspace}
\usepackage{color}
\usepackage{array}

\newcommand{\phii}{Institute of Physics II, University of Cologne, 50937 Cologne, Germany}
\newcommand{\geo}{Institute for Geology and Mineralogy, University of Cologne, 50674 Cologne, Germany}

\newcommand{\rucl}{$\alpha$-RuCl$_3$\xspace}

\newcommand{\aug}{Experimental Physics V, Center for Electronic Correlations and Magnetism,
                                University of Augsburg, 86159 Augsburg, Germany}
\newcommand{\hmfl}{High Field Magnet Laboratory (HFML - EMFL), Radboud University, 6525 ED
                                 Nijmegen, The Netherlands}

\newcommand{\frankfurt}{Institut f\"ur Theoretische Physik, Goethe-Universit\"at Frankfurt, 60438 Frankfurt am Main, Germany}

\newcommand{\thp}{Institute for Theoretical Physics, University of Cologne, 50937 Cologne, Germany}

\newcommand{\Korea}{Department of Physics, Chung-Ang University, Seoul 06974, Republic of Korea}

\newcommand{\HZDR}{Institute of Radiation Physics, Helmholtz Zentrum Dresden-Rossendorf, 01328 Dresden, Germany}

\begin{document}
\title{High-Field Quantum Disordered State in \rucl: Spin Flips, Bound States, and a Multi-Particle Continuum}
\author{A.~Sahasrabudhe}
\affiliation{\phii}
\author{D.~A.~S.~Kaib}
\affiliation{\frankfurt}
\author{S.~Reschke}
\affiliation{\aug}
\author{R.~German}
\affiliation{\phii}
\author{T.~C.~Koethe}
\affiliation{\phii}
\author{J.~Buhot}
\affiliation{\hmfl}
\author{D.~Kamenskyi}
\affiliation{\hmfl}
\author{C.~Hickey}
\affiliation{\thp}
\author{P.~Becker}
\affiliation{\geo}
\author{V.~Tsurkan}
\affiliation{\aug}
\affiliation{Institute of Applied Physics, MD2028 Chisinau, Republic of Moldova}
\author{A.~Loidl}
\affiliation{\aug}
\author{S.~H.~Do}
\affiliation{\Korea}
\author{K.~Y.~Choi}
\affiliation{\Korea}
\author{M.~Gr\"uninger}
\affiliation{\phii}
\author{S.~M.~Winter}
\affiliation{\frankfurt}
\author{Zhe~Wang}
\affiliation{\phii}
\affiliation{\HZDR}
\author{R.~Valent{\'\i}}
\affiliation{\frankfurt}
\author{P.~H.~M.~van~Loosdrecht}
\affiliation{\phii}

\date{\today}
\begin{abstract}
Layered \rucl has been discussed as a proximate Kitaev spin liquid compound.
Raman and THz spectroscopy of magnetic excitations 
confirm 
that the low-temperature antiferromagnetic ordered phase 
features
a broad Raman continuum, together with two magnon-like excitations at 2.7 and 3.6 meV, respectively. 
The continuum strength is maximized as long-range order is suppressed by an external magnetic field. 
The state above the field-induced quantum phase transition around 7.5~T is characterized by a gapped  multi-particle continuum out of which a two-particle bound state emerges, together with a well-defined single-particle excitation at lower energy. 
Exact diagonalization calculations demonstrate that Kitaev and off-diagonal exchange terms in the Fleury-Loudon operator are crucial for the occurrence of these features in the Raman spectra. 
 Our study firmly establishes the partially-polarized quantum disordered character of the high-field phase. 
\end{abstract}
\maketitle
%
%
Integrable models are of great interest, as the relevant physics of such models can be obtained in an exact sense. A canonical example is the fractionalization of excitations in the one-dimensional spin-1/2 Heisenberg-Ising chain.
While a single spin-flip changes the total spin by 1, the elementary
excitations of this model are spinons, which carry a fractional quantum number of spin 1/2 \cite{Faddeev81}. Spin fractionalization also occurs in two-dimensional integrable models, with a particularly prominent example being the spin-1/2 Kitaev honeycomb model composed of bond-dependent Ising interactions \cite{kitae06}. Its exact solution results in an exotic quantum spin-liquid (QSL) ground state, where spin-flip excitations fractionalize into gapless Majorana fermions and gapped flux excitations \cite{kitae06,trebs17,winter2017models,herma18,takag19}. 

Candidate materials for realizing the fascinating
physics of the Kitaev model feature Ir$^{4+}$ or Ru$^{3+}$ ions with spin-orbit entangled $j=1/2$ moments \cite{jacke09}.
The presence
of dominant ferromagnetic Kitaev interactions is
well established in \textit{A}$_2$IrO$_3$ (\textit{A} = Na or Li) and  \rucl \cite{baner17,trebs17,winter2017models,herma18,takag19}. However, they appear simultaneously with subleading isotropic Heisenberg and anisotropic off-diagonal exchange interactions \cite{trebs17,winter2017models,herma18,takag19}, resulting in long-range order below a finite temperature $T_\text{N}$, hampering experimental observation of the Kitaev QSL. 

There remain though promising alternate avenues to observing the sought-after QSL physics. On the one hand, by going to elevated temperatures above $T_\text{N}$, various spectroscopic studies of spin dynamics have found features reminiscent of the Kitaev QSL in \rucl \cite{sandi15,baner16,do17,wang17,wang19,resch19}.
On the other hand, one may suppress the long-range order by tuning external parameters \cite{johns15, leahy17, sears17, wolte17, baek17,winte18,jansa18,wang17,ponom17,zheng17,hentr18,baner18,lampe18b,wellm18,kasah18b,balz19}, e.g.\ magnetic field, in order to hopefully detect the QSL as a field-induced phase, before reaching the partially polarized quantum disordered state (QDS), that is smoothly connected to the fully field-polarized limit (due to the lack of SU(2) symmetry, the polarized state is only approached asymptotically with increasing field strength). 

In \rucl, 
a variety of experimental studies have reported a single field-induced quantum phase transition with an in-plane critical field of $B_c = 6-8$~T \cite{johns15, leahy17, sears17, wolte17, baek17,jansa18,wang17,ponom17,zheng17,hentr18,baner18}. The nature of the subsequent field-induced phase has been under strong scrutiny, with conflicting reports over whether it is gapped or gapless and whether it is the sought-after QSL or the partially field-polarized state \cite{johns15, leahy17, sears17, wolte17, baek17,jansa18,wang17,ponom17,zheng17,winte18,hentr18,baner18}. We also note that in some more recent studies, additional field-induced phase transitions were reported \cite{lampe18b,wellm18}, within the narrow field range $6-10$~T. An intermediate region with quantized thermal Hall conductivity has further been reported \cite{kasah18b}.

Raman scattering has been suggested as being particularly powerful in revealing signatures of spin fractionalization of the Kitaev QSL as it naturally probes the emergent Majorana fermion excitations \cite{knoll14}. 
In \rucl, the experimentally observed broad Raman continuum at zero field was taken as evidence for spin fractionalization \cite{knoll14,sandi15}, while in particular its temperature dependence was claimed to reflect the character of fermionic excitations \cite{sandi15,nasu16,wang19}.
The Raman response of the possible field-induced phases so far remained unexplored.

In this work, we study the magnetic excitations of \rucl as a function of in-plane magnetic field up to 33~T using Raman and THz spectroscopy.  
The Raman continuum exhibits maximum intensity as soon as the long-range zig-zag (ZZ) antiferromagnetic order is suppressed at $B_c=7.5$~T. 
In the high-field limit, we observe a gapped continuum of multi-particle excitations. Below this continuum, both Raman and THz spectroscopy reveal a two-particle bound state as well as a sharp single-particle excitation. These features are most clearly resolved in the
high-field regime above 15~T, yielding a clear picture of the properties 
of the high-field phase.
At intermediate fields near $B_c$, the features start to overlap in energy, which 
partially explains previous difficulties in understanding the field-induced phase. The access to the high-field limit further allows us to resolve a long-standing controversy regarding anomalously steep slopes \cite{wang17,ponom17,winte18} of the excitation modes at intermediate fields. 
Our numerical calculations essentially capture the field-dependent features, and highlight the crucial role of the Kitaev exchange and off-diagonal interactions in observing Raman features that are absent in conventional magnets.

High-quality \rucl crystals for our study were prepared by vacuum sublimation. They exhibit a sharp magnetic phase
transition at $T_\text{N}=6.5$~K \cite{Widmann19, do17}.
High-field Raman back-scattering and THz-transmission experiments on samples from Augsburg \cite{Widmann19,resch18} and Seoul \cite{do17,wang17} were performed in Bitter magnets for in-plane fields up to 30 and 33~T, respectively, at a cryostat base temperature of 1.7 and 2~K.
Both experiments were carried out in a Voigt geometry, i.e.\ $\mathbf{B} \perp \mathbf{k}$, with the incident wave vector $\mathbf{k}$ perpendicular to the sample's hexagonal {\it ab}-plane.
The field was oriented nearly perpendicular (within 10$^\circ$) to the nearest-neighbor Ru-Ru bonds,
while the in-plane field orientation was not determined for the THz measurements.
For Raman spectroscopy, circularly polarized (\emph{L} for left and \emph{R} for right) light with a wavelength of 532 nm was focused into a spot of about 2~$\mu$m on freshly cleaved samples.
THz transmission spectra were recorded on samples with typical \textit{ab} surface of $3\times3$ mm$^2$ and thickness of 1~mm using a Fourier-transform spectrometer Bruker IFS-113v, with a mercury lamp as source and a silicon bolometer as detector.

\begin{figure}
\includegraphics[width=0.95\linewidth]{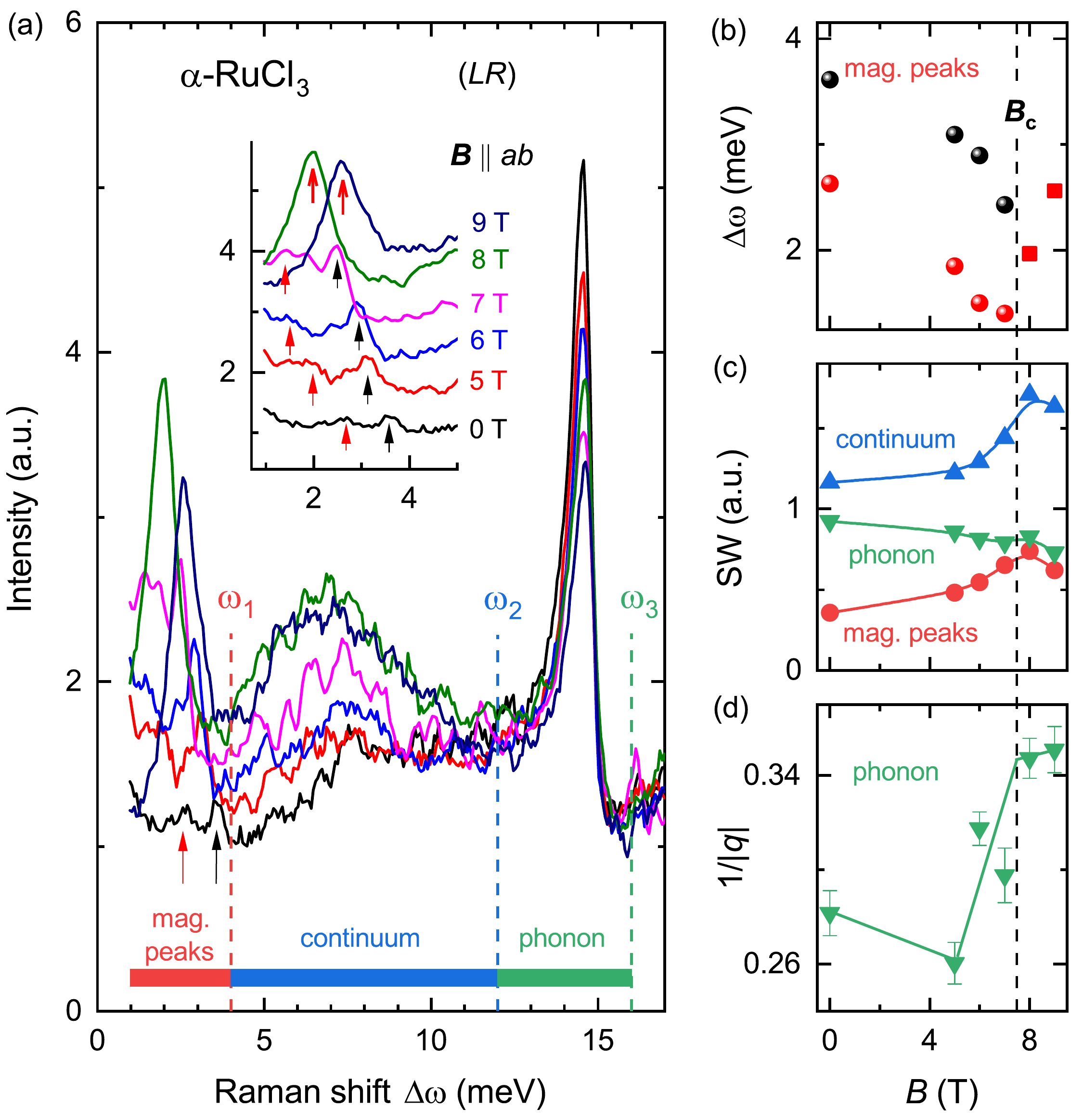}
     \caption{
     (a) Circularly-polarized (LR) Raman spectra recorded at 1.7~K with low incident laser power of 10~$\mu$W in fields up to 9~T. Inset: Magnetic excitations are resolved at low energies, as indicated by the arrows. The spectra of different fields are shifted upward by a constant for clarity.
     (b)~Energy transfer of the magnetic excitations in (a) is shown as a function of field.
     (c)~Spectral weight for different frequency ranges, as indicated by the dashed lines in (a).
     (d)~Fano parameter $1 /\vert q \vert$ for the 15~meV phonon. The solid line is guide to the eye.
     The dashed line in (b)-(d) indicates the critical field $B_c=7.5$~T.
} 
     \label{lowfield}
\end{figure}

\begin{figure}
\includegraphics[width=0.9\linewidth]{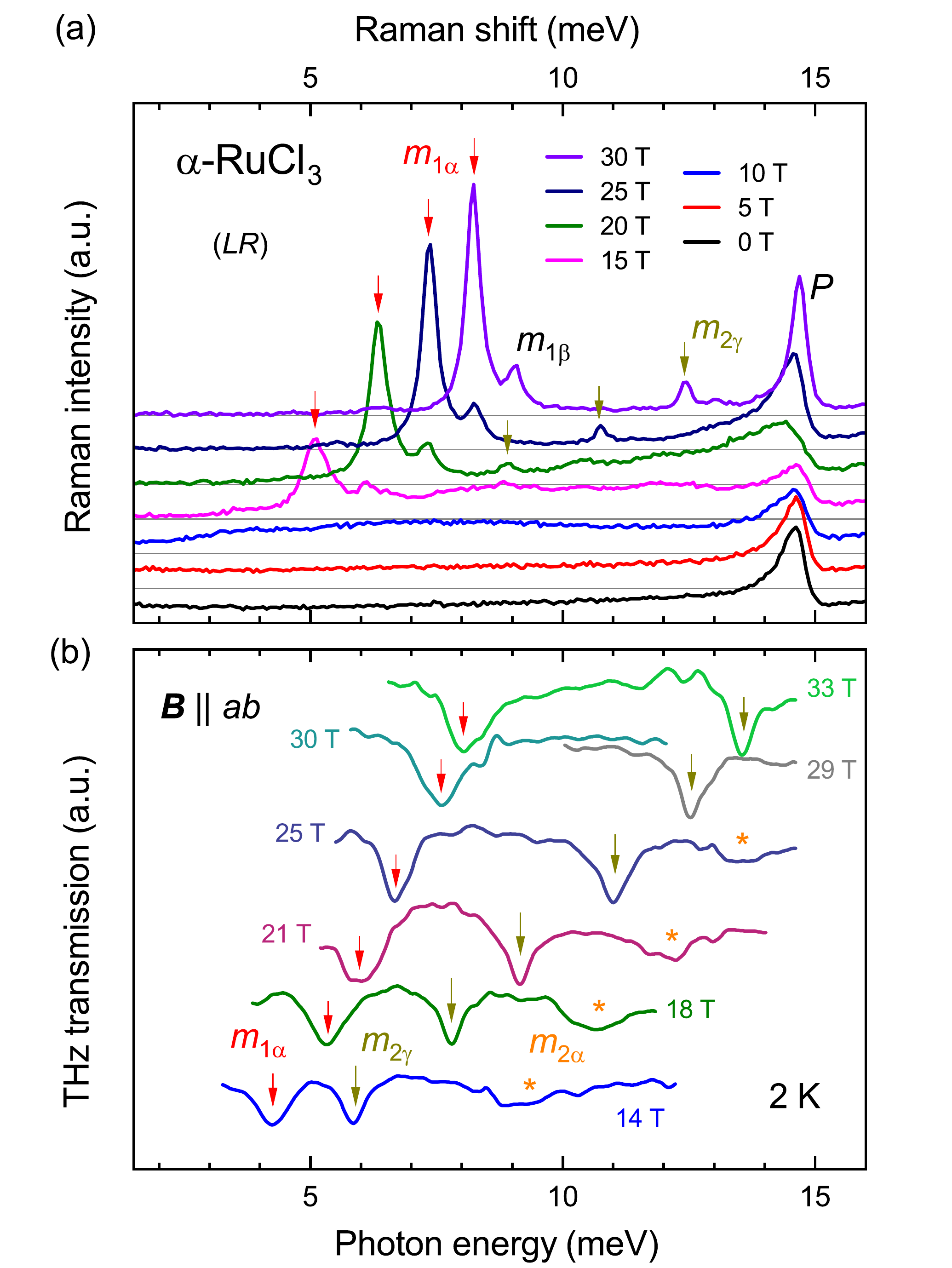}
     \caption{ High-field (a) 100~$\mu$W \emph{LR}-Raman and (b) unpolarized THz-transmission spectra. A variety of field-dependent features are indicated (see text for details). \emph{P} denotes phonon.
     }
     \label{highfield}
\end{figure}

To minimize heating effects due to the incident beam, we applied a very low laser power of 10~$\mu$W for the Raman spectroscopy of the low-field ordered phase.
Figure~\ref{lowfield}(a) presents the Raman spectra in the \textit{LR}-circularly polarized channel in magnetic fields of up to 9~T. While the zero-field Raman spectrum is dominated by a broad scattering continuum in accordance with previous Raman studies \cite{sandi15,glama17,wang19}, we can also discern two peaks of magnetic excitations in the low-energy regime at 2.7 and 3.6~meV, respectively, as indicated by the arrows in Fig.~\ref{lowfield}(a) and its inset, which shift systematically to lower energies with increasing magnetic field up to 7~T [Fig.~\ref{lowfield}(b)].
These two modes, in energy and field dependence comparable to those reported in previous neutron scattering \cite{baner16}, ESR \cite{ponom17}, and THz spectroscopic studies \cite{wang17,littl17,shi2018absorption,little18}, correspond to the single-magnon excitations at the $\Gamma$ point of the ZZ ordered state. While this observation clearly indicates that our sample temperature stayed well below $T_\text{N}$, we found that increasing the laser power to 100~$\mu$W is sufficient to erase these two modes from the Raman spectra [see Fig.~\ref{highfield}(a)], which explains the absence of these modes in the previous Raman studies.
Above $B_c=7.5$~T, only a single peak dominates the low-energy Raman response [Fig.~\ref{lowfield}(a)]. In particular, for 8~T,
where evidence for a separate intermediate phase has been reported in recent thermodynamic measurements \cite{kasah18b,balz19}, we find a sharp peak $m_{1\alpha}$ at 2~meV in the Raman spectrum. Consistent with reported THz results \cite{wang17,ponom17}, this mode hardens continuously in higher fields above $B_c$ (Fig.~\ref{highfield}), without any obvious features indicating a second field-induced phase transition. 

At higher energies, the zero-field Raman response presents a set of phonon modes, in accordance with previous Raman studies \cite{sandi15,glama17,wang19}. The lowest phonon at 15~meV exhibits an asymmetric Fano lineshape with a pronounced field dependence [see Fig.~\ref{highfield}(a)]. 
The Fano lineshape reflects interference of the phonons with an underlying continuum.
The lineshape asymmetry quantified by the Fano parameter $q$ is represented as $1/\vert q \vert$ in Fig.~\ref{lowfield}(d). Approaching the critical field from below, $1/\vert q \vert$ displays an abrupt increase,
indicating that the interference is strongly enhanced at $B_c$. This also indicates that the 15 meV phonon modulating the Ru-Ru bond \cite{glama17} is strongly coupled to the magnetic degrees of freedom.

In addition to the high-energy phonons and the low-energy magnon-like excitations, the zero-field spectrum is marked by an extended continuum \cite{sandi15,wang17,glama17,wang19,baner17,do17,resch19}, that shows a broad intensity maximum in the mid-energy region between 4 and 12~meV. 
 Figure~\ref{lowfield}(c) displays the integrated spectral weight for three different energy regions [indicated by dashed lines in Fig.~\ref{lowfield}(a)] as a function of field strength. While the spectral weight of the 15~meV phonon is reduced at higher fields, both the low-energy excitations and the mid-frequency continuum are clearly enhanced upon approaching the field-induced quantum phase transition at $B_c=7.5$~T.

To obtain a comprehensive picture of the field-induced phase above $B_c$, we carried out both Raman and THz spectroscopic measurements, the results of which are presented in Fig.~\ref{highfield}.
With an increased laser power of 100~$\mu$W, the signal-to-noise ratio of the Raman spectra is improved, allowing us to resolve the high-field features.
At the same time, due to additional heating effects, the low-energy magnons that were present in Fig.~\ref{lowfield}(a) cannot be resolved here,
and the $m_{1\alpha}$ mode seen sharply at 8~T is also weakened, but becomes stronger and easily recognized above 10~T. At 15~T, the $m_{1\alpha}$ mode is observed at 5.1~meV and accompanied by a satellite peak $m_{1\beta}$ at 6.1~meV, both of which become sharper and shift to higher energy with increasing fields. While the field dependence of the $m_{1\alpha}$ mode is consistently seen by the THz spectroscopy data [Fig.~\ref{highfield}(b)], the satellite peak is hardly resolvable.
It is worth noting that the $m_{1\alpha}$ peak position in the THz spectra is located slightly lower, about 0.6~meV, than in the corresponding Raman spectra, which comparable to in-plane anisotropy \cite{little18} can be due to uncertainty in our in-plane field orientation.

Another sharp feature in the Raman spectra is the observation of mode $m_{2\gamma}$ at higher energy, which is located at 12.4~meV at 30~T 
and which shows a steeper increase in energy with field, as compared with the dominant $m_{1\alpha}$ mode. Consistent energy and field dependence is also found in the THz spectra, but the relative intensity $I_{m_{2\gamma}}/I_{m_{1\alpha}}$ is much stronger in THz [Fig.~\ref{highfield}(b)].

\begin{figure}
\includegraphics[width=0.95\linewidth]{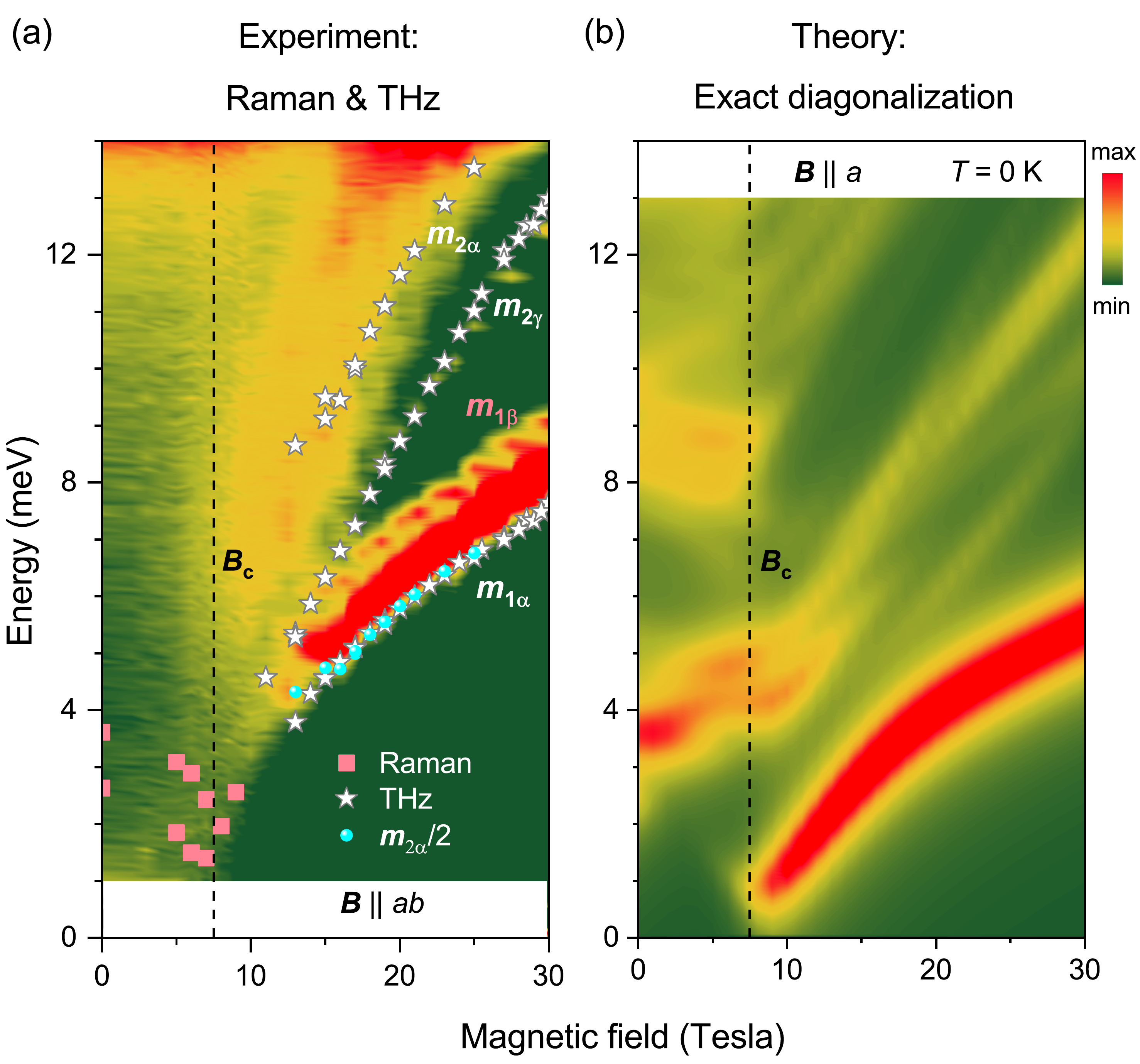}
     \caption{
     (a) Contour plot of experimental 100~$\mu$W Raman intensity of \rucl as a function of in-plane field [from Fig.~\ref{highfield}(a)]. Peak positions obtained in the low-field 10~$\mu$W Raman spectra ($B\le 8$~T) [Fig.~\ref{lowfield}(b)] and in the high-field THz spectra ($B\ge 10$~T) [Fig.~\ref{highfield}(b)] are shown by symbols for comparison. 
     (b) Theoretical $T=0$ Raman response within the Fleury-Loudon approximation \cite{fleur68} of a $C_3$-broken model for $\mathbf B \parallel a$ assuming $g_{ab}=2.3$ \cite{supplemental}. 
     }
     \label{overview}
\end{figure}

A unique feature present in the high-field Raman response is the underlying continuum of excitations. In contrast to the spectra below $B_c$, one can clearly see a gap opening at $B_c$ with the lower bound of the continuum shifting towards higher energy upon increasing field. For instance, at 15~T, the spectral weight below 3~meV is almost fully depleted, just below the $m_{1\alpha}$ mode, signaling a gap opening.
 The lower bound of the continuum shifts with a similarly steep slope as $m_{2\gamma}$ to higher energies, approximately twice of that of $m_{1\alpha}$. These features are better illustrated in the color plot of the Raman intensity in Fig.~\ref{overview}(a). Usually, a featureless continuum is rare to resolve in THz spectroscopy, but as marked by the asterisk in Fig.~\ref{highfield}(b), we can track a broad transmission minimum $m_{2\alpha}$ as a function of field, in agreement with that of the Raman continuum [Fig.~\ref{overview}(a)]. Thus, the THz $m_{2\alpha}$ band captures the broad maximum of the underlying continuum with its strong field dependency.

The fact that the modes $m_{1\alpha}$ and $m_{2\gamma}$ are observed both in the THz response and in Raman aids significantly in their identification. Based on comparison with previous linear-polarized lower-field THz studies \cite{wang17,winte18}, we identify the $m_{1\alpha}$ mode as a single-particle excitation expected in the QDS, which is well described by linear spin-wave theory as a single-magnon only at sufficiently high fields. 
Going to stronger fields we observe that the higher-energy continuum exhibits a somewhat broad maximum $m_{2\alpha}$, which has approximately twice the energy of $m_{1\alpha}$, i.e.\ $m_{2\alpha}=2m_{1\alpha}$ [Fig.~\ref{overview}(a)]. This identifies the lower band near $m_{2\alpha}$ of the multi-particle continuum to consist predominantly of two-particle excitations. 
 However an additional observation enabled by going to higher fields is the clear separation of an apparent two-particle bound state $m_{2\gamma}$ slightly below the multi-particle continuum and well above $m_{1\alpha}$. The two-particle nature of $m_{2\gamma}$ is consistent with the polarization dependence in THz \cite{wang17,winte18}. Finally, the field dependence of the satellite peak $m_{1\beta}$ follows that of $m_{1\alpha}$ with a nearly field-independent energy difference of about 1~meV, which suggests that  $m_{1\beta}$ could be due to multiparticle scattering or interlayer coupling. We note that a recent neutron scattering study revealed out-of-plane dispersion of the single-particle excitations with a bandwidth of around 1~meV \cite{balz19}. 

The observation of the single-particle mode $m_{1\alpha}$ in Raman spectroscopy may appear surprising at first. Typically, via the so-called Fleury-Loudon scattering processes \cite{fleur68}, the magnetic Raman response is dominated by two-magnon excitations. It is thus highly unusual to sharply resolve magnetic single-magnon excitations in Raman spectroscopy, such as $m_{1\alpha}$ at high field. To see how it can occur here, consider the infinite-field polarized limit $|\mathbf{B}|\to \infty$, in which case the exact ground state is an eigenstate of $S_\text{tot}^\mu$, with $\mu$ being the direction of $\mathbf{g}\cdot \mathbf{B}$. In this case, terms in the Fleury-Loudon scattering operator like $S_i^\mu S_j^\nu$, with $\mu\neq\nu$, may create single spin-flip ($|\Delta S|=1$) excitations (which correspond to single-magnons in the polarized limit). Indeed, such terms naturally arise in the presence of Kitaev and off-diagonal exchanges, making \rucl a striking example of this unusual Raman response.

An important quantity of interest is the slope of the excitation energy  of the $m_{1\alpha}$ mode as a function of field, $g^\ast\equiv\frac{\mathrm dE_{m_{1\alpha}}}{\mu_{\text B}\mathrm dB}$. In the large-field limit, one expects this to approach $g^\ast  \vert_{B\to\infty} = g_{ab}| \Delta S |$. Since the intrinsic Land\'e in-plane $g$-value is constrained to $2 \le g_{ab}\le 2.8$ %
\cite{chaloupka2016anisotropy,yadav16,winter2017models}, the asymptotic slope provides information regarding $g_{ab}$ and the character of the excitation. In this regard, the reported slopes of $g^\ast \gtrsim 8$ for measurements near the critical field ($B \gtrsim B_c$) \cite{wang17,ponom17} have been 
discussed as evidence for either fractionalized excitations of a field-induced QSL state, an enormous in-plane g-value $g_{ ab}\approx 10$, or an apparent multi-particle bound-state character of mode $m_{1\alpha}$.  
However, the direct comparison of the measured slopes with the polarized limit has two major caveats: (i) Level repulsion between $m_{1\alpha}$ and the continuum may significantly increase the slope in the vicinity of $B_c$, and (ii) since the anisotropic couplings produce an effective easy-plane anisotropy in the QDS \cite{janssen2017magnetization,riedl2019torque}, the slope of the single-particle excitation is expected to approach $g_{ab}$ only asymptotically 
 (see Ref.~\onlinecite{kittel}). 
 Indeed we find that $g^\ast$ drops from 8 at around 10~T \cite{wang17} continuously to about~3 at 30~T.  By analyzing the data over the full field range we extract an asymptotic $ g^\ast  \vert_{B\to\infty} = 2.51 \pm 0.18 $ \cite{supplemental} which confirms $m_{1\alpha}$ as the mode that evolves into the $|\Delta S|=1$ spin-flip excitation in the infinite-field limit.

To complement the experimental results, we perform exact diagonalization calculations of suitable realistic models on a $24$-site cluster \cite{supplemental}, with parameters based on \textit{ab-initio} studies \cite{winte16,yadav16,kim16,wwang17,Hou17}. The Raman intensity $I(\omega)=\int \mathrm dt\, e^{-i\omega t}\langle \mathcal F(t) \mathcal F(0) \rangle$ is evaluated using the Fleury-Loudon approach \cite{fleur68}, with the scattering operator
$\mathcal F \propto \sum_{ij}\, \mathbf S_i \cdot \hat{\mathbf J}_{ij} \cdot \mathbf S_j \, (\boldsymbol \delta_{ij} \cdot \mathbf E_\text{in})(\boldsymbol \delta_{ij} \cdot \mathbf E_\text{out}^*)$
where $\hat{\mathbf J}_{ij}$ contains the generic couplings between $\mathbf S_i$ and $\mathbf S_j$, $\boldsymbol \delta_{ij}$ is the distance vector between sites $i$ and $j$, and $\mathbf E_\text{in}$ ($\mathbf E_\text{out}$) is the direction of the electric field of the incident (outgoing) light. We find that the essential features of the measured Raman response can be captured this way in realistic $(JK\Gamma J_3)$ models, for example the model of Ref.~\onlinecite{winte17} (see Supplemental Material \cite{supplemental}). However not all models capture the strong field dependence of $g^\ast$. Figure~\ref{overview}(b) shows the results for an adjusted model which does capture this, namely 
$(J_1, K_1^{x/y}, K_1^z, \Gamma_1, J_3) =  (-0.5,\,-7.5,\,-5,\,2.5,\,0.5)\operatorname\cdot1.5\text{ meV}$. The employed breaking of $C_3$ symmetry ($K_1^x=K_1^y\neq K_1^z$) is motivated by \textit{ab-initio} calculations \cite{winte16} for the $C2/m$ structure of \rucl\ \cite{johns15,cao16}. 
As observed in the experiment [Fig.~\ref{overview}(a)], a continuum
extends throughout a wide energy range below $B_c$ in the numerical results [Fig.~\ref{overview}(b)], while for $B>B_c$, the continuum rises in energy. Above $B_c$, a strong, sharp mode emerges at lowest energies, which shifts to higher energy and becomes the most intense excitation at high fields. This mode follows the field-induced excitation-gap opening, corresponding to the observed $m_{1\alpha}$ mode in Fig.~\ref{overview}(a). We have further studied separately the contributions of the Heisenberg and the off-diagonal exchange terms \cite{supplemental}, and confirmed that the Kitaev and the off-diagonal exchanges in $\mathcal F$ are responsible for the observation of the single-particle excitation in the Raman response.

In conclusion, our study shows that the high-field phase 
in \rucl, which emerges through the suppression of 
antiferromagnetic order, is neither a quantum 
spin-liquid state nor a fully field-polarized state, but rather a quantum disordered state with partial field alignment of the 
spin-orbital moments. No clear evidence is found for an intermediate phase around 7.5~T in our experiments. The 
high-field phase is spectroscopically characterized by a strong and sharp single-particle excitation and a continuum of multi-particle nature, out of which a well-defined two-particle bound state
emerges, best visible at higher fields. A comparison 
of experimental and exact diagonalization results clearly demonstrates the importance of Kitaev and off-diagonal interactions in 
\rucl. Based on the current study one expects similar 
high-field physics in other Kitaev candidate materials such as the iridates.

\begin{acknowledgements}
We acknowledge helpful discussions with A.~L.~Chernyshev, J.~Knolle, S.~Trebst, and P.~A.~Maksimov. 
This work is partially supported by the DFG (German Research Foundation) via the project No.\ 277146847 - CRC 1238: Control and Dynamics of Quantum Materials
, via the project No.\ 107745057 - TRR 180:  From Electronic Correlations to Functionality (Subproject F5), and via the project No.\ VA117/15-1: Probing the nature of excitations in spin-orbit-coupled materials from first principles. 
We acknowledge the support of the HFML, member of the European Magnetic Field Laboratory (EMFL).
\end{acknowledgements}

\end{document}